\documentclass[a4paper,10pt]{article}
\usepackage[margin=1in]{geometry}
\usepackage{authblk}
\usepackage{caption}

\usepackage{graphicx}
\usepackage{color}
\usepackage{hyperref}
\usepackage{amsmath}
\usepackage{pdflscape}

\newcommand*\samethanks[1][\value{footnote}]{\footnotemark[#1]}

\begin{document}
\title{Inequality is rising where social network segregation interacts with urban topology}

\author[a,b]{Gerg\H{o} T\'oth\thanks{G.T. and J.W. contributed equally to this work.}}
\author[c]{Johannes Wachs\samethanks} 
\author[d,e]{Riccardo Di Clemente}
\author[f]{\'Akos Jakobi}
\author[a,g,i]{Bence S\'agv\'ari}
\author[h]{J\'anos Kert\'esz}
\author[a,i]{Bal\'azs Lengyel\thanks{Corresponding author: E-mail: lengyel.balazs@krtk.mta.hu}}

\affil[a]{\footnotesize Agglomeration and Social Networks Lend\"ulet Research Group, Hungarian Academy of Sciences}
\affil[b]{Spatial Dynamics Lab, University College Dublin}
\affil[c]{Chair of Computational Social Sciences and Humanities, RWTH Aachen University}
\affil[d]{Centre for Advanced Spatial Analysis, University College London}
\affil[e]{Department of Computer Science, University of Exeter}
\affil[f]{Department of Regional Science, E\"otv\"os Lor\'and University}
\affil[g]{Center for Social Sciences, Hungarian Academy of Sciences}
\affil[h]{Department of Network and Data Science, Central European University}
\affil[i]{International Business School Budapest}

\maketitle

\begin{abstract}
Social networks amplify inequalities due to fundamental mechanisms of social tie formation such as homophily and triadic closure. These forces sharpen social segregation reflected in network fragmentation. Yet, little is known about what structural factors facilitate fragmentation. In this paper we use big data from a widely-used online social network to demonstrate that there is a significant relationship between social network fragmentation and income inequality in cities and towns. We find that the organization of the physical urban space has a stronger relationship with fragmentation than unequal access to education, political segregation, or the presence of ethnic and religious minorities. Fragmentation of social networks is significantly higher in towns in which residential neighborhoods are divided by physical barriers such as rivers and railroads and are relatively distant from the center of town. Towns in which amenities are spatially concentrated are also typically more socially segregated. These relationships suggest how urban planning may be a useful point of intervention to mitigate inequalities in the long run. 
\end{abstract}

\section*{Introduction}

Wealth and wage inequalities are growing \cite{piketty2015capital}, slowing development, economic growth, and technological progress \cite{kawachi1997social, alesina1994distributive, hartmann2017linking} while fostering radicalization and the advance of political populism \cite{abadie2006poverty, inglehart2016trump}.
These disparities are deeply rooted in history; unequal access to education, technology, and public services are self-reinforcing mechanisms by which economic inequality compounds across generations \cite{kuznets1955economic, lucas2001effectively}.

Geography is both an important source and marker of economic inequalities. A stylized fact suggest that home location describes much of individuals' economic potential and access to opportunities through education \cite{chetty2014land}. A consequent divergence of outcomes across neighborhoods is observed even within relatively small geographical units such as cities and towns \cite{glaeser2009inequality,sampson2008moving}.

More recently, high-resolution cellphone and social media data has enabled researchers to analyze the relationship between social network structure and the overall wealth of cities \cite{eagle2010network,norbutas2018network}. Their findings support previous theoretical claims that social networks offer access to resources and economic opportunities \cite{granovetter1985economic}. It is also thought that social networks have an important role in the unequal distribution of resources \cite{freeman1978segregation,dimaggio2012network}. For example, the micro-level tendencies thought to explain the formation of social ties such as homophily, the tendency of similar individuals to become friends \cite{mcpherson2001birds}, and triadic closure, the tendency of friends of friends to become friends \cite{bianconi2014triadic}, can result in social segregation at the macro scale \cite{stadtfeld2018micro}. This kind of macro-scale network topology can lead to economic inequalities between groups if access to resources or information runs through the network \cite{dimaggio2012network}. Yet to our knowledge, big data on social networks have not tested the relationship between social segregation and economic inequality.

In this paper, we analyze a large scale online social network of ca. $2$ Million individuals locate in ca. $500$ towns of Hungary and investigate how the fragmentation of the social network within towns is related to levels and changes of income inequality between years $2011$ and $2016$. While such social network data have been used to relate network analysis structure and overall economic outcomes \cite{eagle2010network,norbutas2018network} our empirical analysis provide novel evidence at large scale about the relationship between social segregation and distributional outcomes, namely inequality.

Why do we observe different levels of social segregation in different cities if universal micro-scale network formation mechanisms are the origins of social segregation? One potential reason is that social interactions are embedded in and constrained by physical space. In cities, individuals sort themselves or are sorted into neighborhoods by income level \cite{reardon2011income} or by communities such as ethnic groups \cite{chodrow2017structure}. Indeed geographic proximity between individuals is a good  predictor of their similarity \cite{calabrese2011interplay}. The urban activities of individuals are been shown to cluster in the city space accordingly with their age, gender and income \cite{Di_Clemente_2018}. Researchers of city science have therefore proposed practical measures to reduce extreme inequalities by improving the access between neighborhoods \cite{batty2013new, brelsford2018toward}. What has not yet been quantified is the extent to which urban topology relates to actually observed social segregation. This is an important gap in the literature as social networks are a primary hypothesized channel by which geographic constraints are related to economic inequalities. Moreover, unlike social networks, which are difficult to change directly via public policy interventions, cities are significantly shaped by urban planning and policy choices made by governments.

To understand how the structure of built environment relates to income inequalities through social relations, we use open source geographic data and develop a composite urban topology index incorporating three measures of urban segregation of towns: 1. the average residential distance from the town center, 2. the degree to which physical barriers divide residential areas, and 3. the extent of spatial concentration of amenities in towns. Each of these indicators and their composite measure are significantly related to social network fragmentation. Using a machine learning approach, we find that these geographic indicators are better predictors of social network fragmentation than other social indicators of segregation.

We find empirical evidence that income inequalities rise more in towns where social networks are fragmented and initial income inequalities are also high. We deploy the urban indices as instrumental variables for social network fragmentation in a model predicting economic inequality. The model shows that our geographic indicators have a significant relationship with economic inequality via their relationship with social network fragmentation.

\section*{Results}

We first investigate the levels and changes of income inequality from 2011 to 2016 in all 474 Hungarian towns with at least 2500 inhabitants. (The capital Budapest, which is an order of magnitude larger than the second largest city, is excluded from the analysis.) We measure income inequality using the Gini index (see Supporting Information 1) and then relate it to our measure of social network fragmentation at the town level.

\begin{figure*}[!ht]
\centering
    \includegraphics[width=\linewidth]{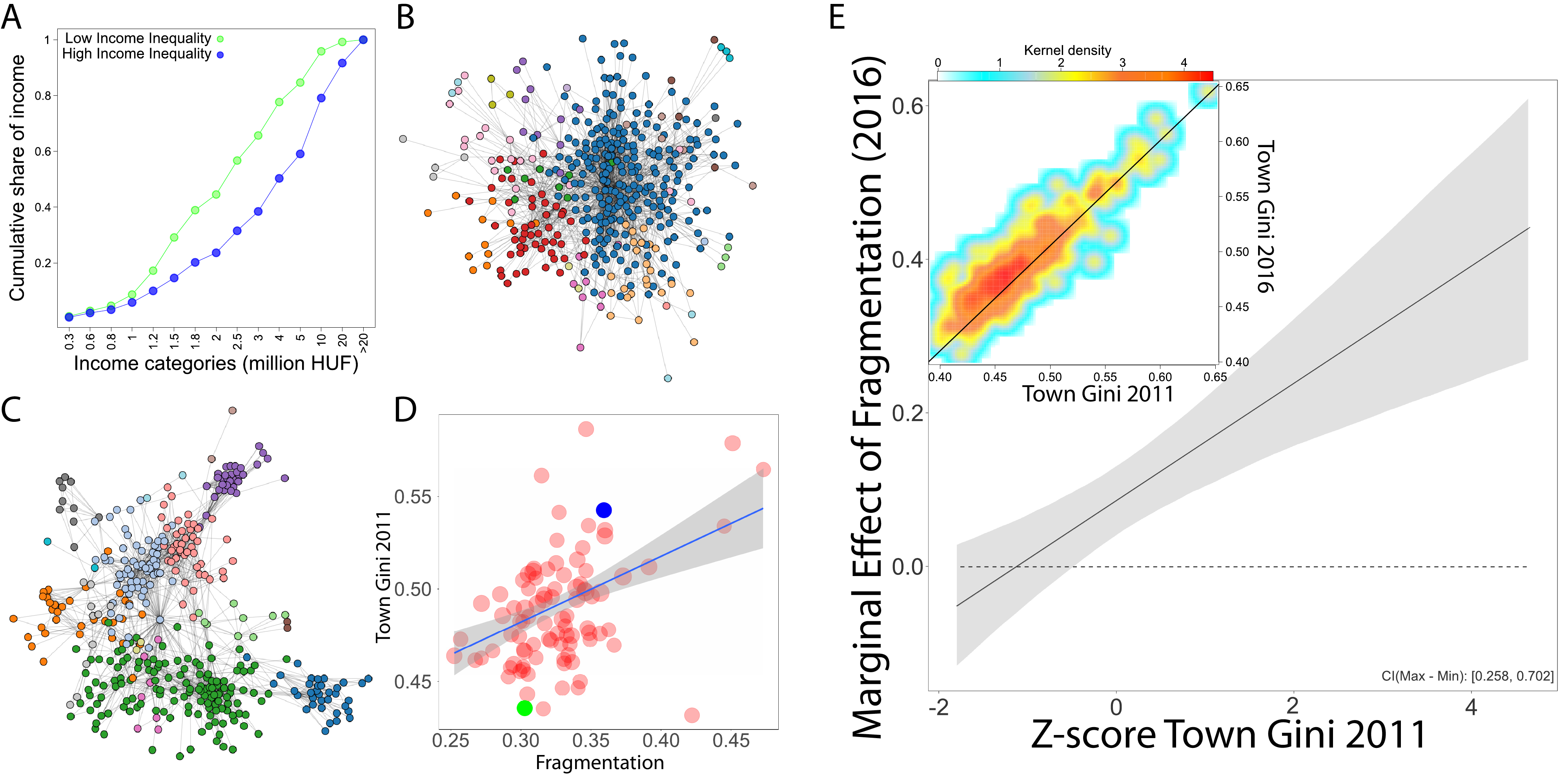}
\caption{\textbf{Income inequality correlates with network fragmentation in towns.} \textbf{(A)} Cumulative distribution of income in a relatively equal town (Ajka, green line) and a relatively unequal one (G\"od\"oll\H{o}, blue line). \textbf{(B)} The social network structure in Ajka, the sample town that has low income inequality. \textbf{(C)} The social network structure in G\"od\"oll\H{o}, the sample town that has high income inequality. \textbf{(D)} Income inequality (measured by the Gini index)  correlates with the fragmentation of social network within the town (Pearson's $r$ = 0.44 for towns larger than 15,000 population). \textbf{(E)} Network fragmentation intensifies income inequality stronger in those towns  where initial inequality is high. $\beta$, the marginal effect of town social network fragmentation $(F_{i})$ on the Gini of the town in 2016 $(G_{i,2016})$, becomes significant around the mean of the Gini in 2011 $(G_{i,2011})$ i.e. at $Z_{G_{i,2011}}=0$. It increases as $G_{i,2011}$ grows. \textit{In the subplot we plot the correlation between town Gini scores in 2011 and 2016 ($G_{i,2011}$ and $G_{i,2016}$.)}}
\label{fig:Fig1}
\end{figure*}

The Hungarian Statistical Office provides binned data on personal income tax filings in each town in our sample (see Materials section).
%\cite{HSO}
As an example, in Figure \ref{fig:Fig1}A we compare the cumulative distribution of gross income across these bins in 2011 for a low (Ajka, in green) and high inequality town (G\"od\"oll\H{o} in blue), both having around 30,000 inhabitants. The green cumulative distribution is above the blue one, indicating that there is a more moderate divide between low- and high income individuals in the former. Denoting the Gini index of town $i$ in year $t$ by $G_{i,t}$, this means $G_{\textrm{Ajka},2011}<G_{\textrm{G\"od\"oll\H{o}},2011}$.

To capture social network structure within towns, we use data retrieved from a Hungarian online social network (OSN) named iWiW that was a popular social media platform on which nearly 30\% of the country's population registered. Similar OSN data retrieved from other platforms (eg. the Dutch OSN Hyves and Facebook) have been used to predict average income in geographical areas in the Netherlands and in the USA \cite{norbutas2018network, bailey2018social}. In iWiW, we have access to the location of users at the town level and can analyze more than 300 million friendship ties the users have established by 2011. Previous research demonstrated that administrative and geographical boundaries influence the iWiW network \cite{lengyel2015geographies}; successfully modelled diffusion and churn on the network  \cite{lengyel2018role, lHorincz2019collapse}. The 
representativity of iWiW regarding age categories in towns is illustrated in Supporting Information 2.

When studying social network fragmentation within towns, we consider only those links in iWiW, for which both ends correspond to person in the same town.
We apply the community detection method known as Louvain algorithm \cite{blondel2008fast}.
This method partitions the individuals of the network in town $i$ into groups by optimizing a measure called modularity $Q_{i}$ that compares the density of edges within groups to the density across groups \cite{newman2006modularity}.
Mathematically, 
$$Q_{i}=\sum_{k=1}^{K_i}\Big[ \dfrac{L^{w}_{k}}{L_{i}}-\Big(\dfrac{L_{k}}{L_{i}}\Big)^{2} \Big]$$, where $K_i$ is the number of communities for the partition and $L_{i}$ is the total number of edges in town $i$, $L_{k}$ is the number of edges adjacent to members of community $k$, and $L_{k}^{w}$ is the number of edges within community $k$. 

Because $Q_{i}$ is highly dependent on the size and density of the network, following \cite{sah2017unraveling}, we scale it by the theoretical $Q^{max}_{i}$ that would be achieved if all edges were within the communities. 
The ratio
\begin{equation}
F_{i}=Q_{i}/Q^{max}_{i}
\end{equation}
for the town networks provides a good quantitative characterization of their fragmentation~\cite{wachs2019social}. Here we use the values of fragmentation $F_i$ for the year 2011.

The structure of the social networks in the sample towns Ajka and G\"od\"oll\H{o} is illustrated by randomly selecting 200 nodes from their social networks in Figure \ref{fig:Fig1}. The network in the relatively low inequality town Ajka in Figure \ref{fig:Fig1}B is 
well connected compared to the network of the relatively unequal town G\"od\"oll\H{o} in Figure \ref{fig:Fig1}C that is rather fragmented and falls into loosely connected subgraphs. Figure \ref{fig:Fig1}D illustrates the positive correlation (Pearson's $r$ = 0.29 for all towns) between $G_{i, 2011}$ and $F_{i}$ meaning that income inequalities are generally higher in those towns where the social network is fragmented.

Turning to the dynamics of inequality, the subplot in Figure \ref{fig:Fig1}E illustrates the strong correlation ($r = 0.9$) between the $G_{i, 2011}$ and $G_{i, 2016}$. However, we observe slight increases in the inequality in most towns from an average Gini index of 0.474 in 2011 to an average of 0.484 in 2016 (the shift is significant according to the Mann-Whitney U-test). 

To analyze how network fragmentation is related to this dynamics, we use the following ordinary least-squares (OLS) regression:
$$
G_{i,2016}= \alpha \times G_{i,2011} + \beta \times F_{i} + \gamma \times (G_{i,2011} \times F_{i}) + Z_{i,2011}  + \epsilon
$$
where the coefficient $\gamma$ of the interaction term informs us about the joint effect of inequality and network fragmentation. $Z_{i,2011}$ refers to a matrix of control variables (population density and fraction of iWiW users in total population). Here $\beta$ is the regression coefficient for $F_i$ and the total contribution of network fragmentation to income inequalities can be estimated 
from the partial derivative of $G_{i,2016}$ with respect to $F_{i}$ using the formula $\frac{\partial G_{i,2016}}{\partial F_i} = \beta +\gamma \times G_{i,2011}$. 

Figure \ref{fig:Fig1}E presents the relationship between social network fragmentation and the change of town income inequality between 2011 and 2016. We find that the interaction between inequality in 2011 and fragmentation has a positive and statistically significant relationship with inequality in 2016. This result provide empirical support to the theory that social networks can increase inequalities when individuals sort based on their initial endowments \cite{dimaggio2012network}.

Next we analyze the indirect relationship between the sources of network fragmentation and inequalities through their relationship with network structure. We focus our attention on the topology of urban space since it has been considered as an exogenous predictor of inequalities across neighborhoods \cite{cutler1997ghettos,ananat2011wrong}.

To test the hypothesis that urban topology is related to income inequality via its relationship to social network fragmentation, we apply a two-stage least square (2SLS) regression model on income inequality. This two stage estimation allows us to examine the chain of relationships from urban topology to income inequality via social network fragmentation.

In the first stage of the 2SLS model, we estimate social network fragmentation using the formula: 
\begin{equation}
     {F}_{i}=\delta+\gamma IV_{i}+\delta N_{i}+e_{i}
\end{equation}
where $IV_{i}$, short for instrumental variable, denotes our urban topology indicators to be introduced below, $N_{i}$ is the fraction of the population of a town $i$ on iWiW, and $e_{i}$ is an error term, assumed to be normally distributed. 

The urban structure indicators $IV_{i}$ are created using data from open-source geographic databases. This allows the replication of our measurements in other countries. The following indicators are proposed to quantify three dimensions of spatial segregation, the details of which are described in the section on methods.

\paragraph*{Average Distance from the Center ($ADC$)} Downtown is assumed to be and indeed functions as the major hub for social interaction in most towns and cities \cite{alonso1964location}. Because co-location is important for social tie creation and the probability of links decreases as distance grows \cite{lengyel2015geographies}, distance from the center can influence the structure of social networks in towns. We expect that social network fragmentation is higher in towns where the average distance from the center is large, because distant individuals are less likely to meet.

Although we cannot test a causal effect of $ADC$ on social network fragmentation, we do argue that reverse causality is not likely. City growth is a complex phenomenon depending on land use, regulations, economic attractiveness, and transport \cite{batty2008size}. Hence, it is not likely that the presence of segregated social groups drives town growth and hence increases distances, especially not in the short or medium term.

\paragraph*{Segregation by Physical Barriers ($SPB$)} Both built structures such as major roadways and railroad tracks and natural barriers like rivers are known to facilitate segregation in cities \cite{cutler1997ghettos}. The effect of such barriers is thought to be an exogenous factor facilitating segregation. Because they can be considered exogenous, they have been used as instrumental variables to measure the effect of racial segregation on disparities in income \cite{ananat2011wrong}. This measure encodes the colloquial phenomenon that some neighborhoods are on the ``wrong side of the tracks''. The effects of physical segregation on socio-economic outcomes are remarkably persistent. For example, neighborhoods in the eastern parts of post-industrial British cities have lower incomes today because they were less desirable places to live in the 19-th century when the wind (blowing west to east) concentrated pollution there \cite{heblich2016east}. 

In our specific context, we expect that social networks are more fragmented in towns that are spatially segregated both because physical barriers decrease the probability of face-to-face interaction \cite{lengyel2015geographies} and because they facilitate sorting of new arrivals by providing clear boundaries to neighborhoods \cite{benton2018just}. 

Though we cannot demonstrate a causal effect of $SPB$ on network fragmentation, we believe that reverse causality is highly unlikely due to the large time lag. For example, the backbone of Hungarian transportation infrastructure was designed and built in the 19th and early 20th centuries \cite{f2002varosrendezesi}, after which very few new railroad tracks have been built. 

\begin{figure*}[!t]
\centering
\includegraphics[width=\linewidth]{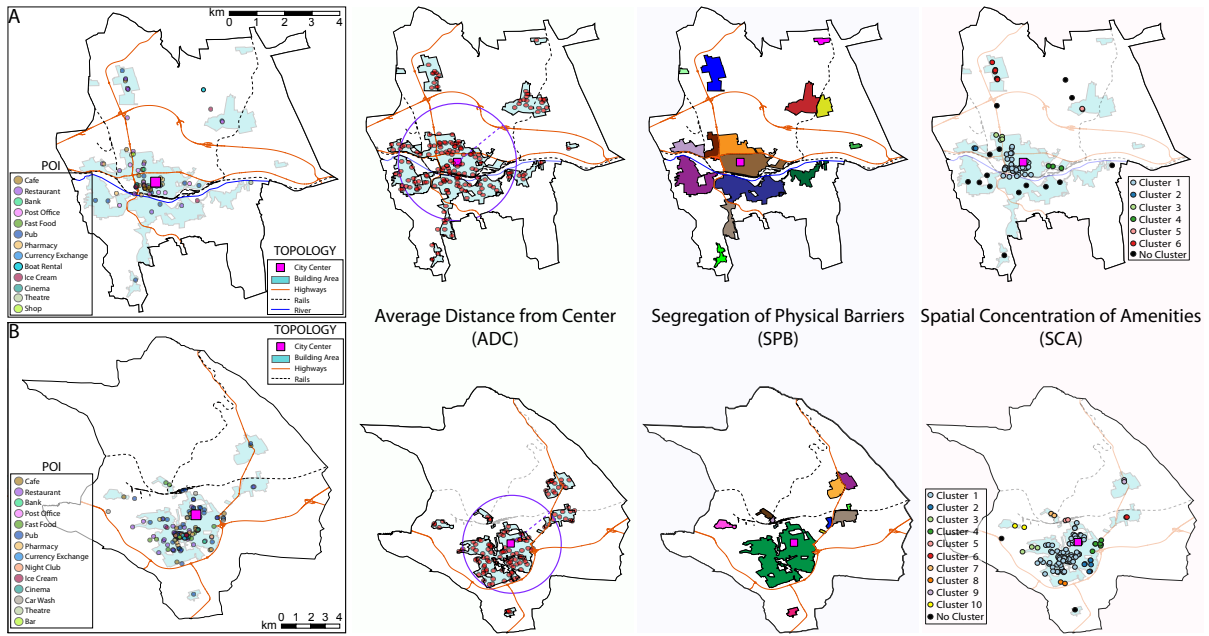}
\caption{\textbf{Urban topology indicators.} A spatially segregated town (Kaposv\'ar) in \textbf{A} has distant fragments of residential zones, increasing the ADC index. It is split into many disconnected components by railways, roads, and rivers, indicated by colors, increasing its SPD index. The scattered meeting points in the town decreases its SCA index. A town with low spatial segregation (Veszpr\'em) in \textbf{B} has relatively compact structure with few distant residential zones and low degree of disconnection by physical barriers. The meeting points are more spatially concentrated in \textbf{B} than in \textbf{A}.}
\label{fig:Fig3}
\end{figure*}

\paragraph*{Spatial Concentration of Amenities ($SCA$)} Individuals go out and interact in places that are not necessarily located downtown \cite{carmona2012public}. The spatial concentration of such amenities is related to the location of rich and poor in cities. Unlike in the US, amenities concentrated in the center of European urban areas have been found to attract the rich to and push the poor from central locations  \cite{brueckner1999central}. Consequently, when amenities are spatially concentrated, residents in peripheral areas of the town might be excluded from majority of social interaction \cite{zhang2019dynamics}. However, the question how this concentration is related to social network fragmentation is still open, since amenities distributed evenly across neighborhoods can also facilitate interaction among peripheral neighbors and increase local cohesion to the detriment of overall connectedness in the town \cite{williams2019great}. 

To understand how the spatial concentration of amenities is related to network fragmentation, we apply point of interest data (POI) that covers restaurants, bars, pharmacies, cinemas etc. To do justice between the two alternative expectations, $SCA$ quantifies the concentration of amenities across its spatial groups defined by a density-based clustering algorithm. This measure takes high value if amenities are concentrated in few spatial clusters and low value if they are evenly scattered across spatial clusters (see description in the Methods section).

We find a significant positive correlation with $SCA$ and network fragmentation ($r=0.253$), which suggests that network segregation is higher in towns where amenities are spatially concentrated. Therefore, we falsify the alternative hypothesis and expect a relation between $SCA$ and inequalities through network fragmentation. However, we cannot rule out reversed causality, since the location of amenities are based on demand that depend directly on local purchasing power.

For the purpose of comparing towns of various sizes, we scale the $ADC$, $SPB$ and $SCA$ measures by the total residential area of towns \cite{ananat2011wrong}. Figure \ref{fig:Fig3} illustrates these three measures in two similar towns, both with around 60,000 inhabitants. Kaposv\'ar in Figure \ref{fig:Fig3}A is spatially segregated: its residential areas have high $ADC$ and are segregated by railroads, rivers, and primary roads. It has a relatively even spatial distribution of amenities. On the other hand, Veszpr\'em in Figure \ref{fig:Fig3}B has a smaller value of $ADC$ is not segregated by physical barriers and amenities are relatively concentrated in the town center. 

To reduce the above discussed dimensions of urban segregation, we create a composite indicator using principal component analysis, 
combining $ADC$, $SPB$ and $SCA$ measures into a single variable. This \textit{Composite Urban Topology Index} ($CUTI$) takes a high value if urban segregation is high in all dimensions (details can be found in Supporting Information 3).

\begin{figure}[!t]
\centering
\includegraphics[width=0.99\linewidth]{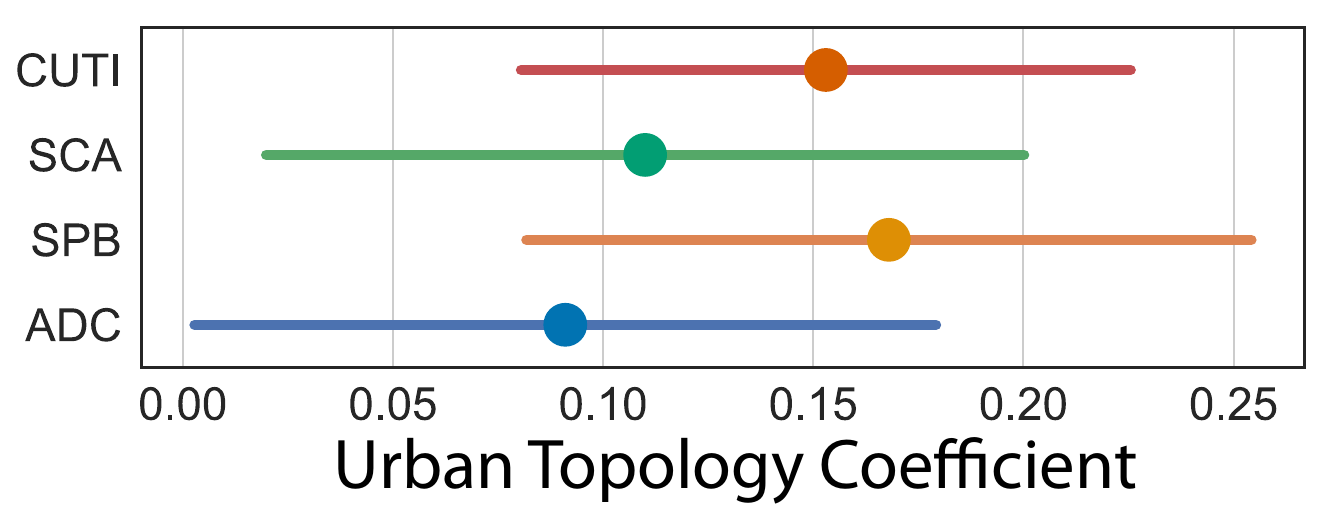}
\caption{\textbf{Social network fragmentation regressed on urban topology indices.} 
Point estimates and $95\%$ confidence interval of urban topology indicators regressed on fragmentation ($F_{i}$). All three urban topology indicators and the composite indicator have a positive, significant relationship with social network fragmentation.
}
\label{fig:Fig4}
\end{figure}

Results presented in Figure \ref{fig:Fig4} confirm a significant positive correlation between network fragmentation $F_{i}$ and all $ADC$, $SPB$ and $SCA$ dimensions of urban segregation. The coefficients of \textit{CUTI} is of comparable size than of the single indicators suggesting that correlation between network fragmentation and various urban topology indicators are robust against the co-dependence of urban topology forms. Description, distribution and correlation of control variables are described in Supporting Information 4. Supporting Information 5 illustrates that every aspect of urban structure outperforms all alternative segregation measures (ethnic, religious, education, political) in predicting network fragmentation by applying a machine learning approach. Supporting Information 6 contains complete regression tables of the first stages of the 2SLS model.

The second stage of the 2SLS estimation follows the formula
\begin{equation}
    G_{i,t}=\alpha+\beta_{1} \hat{F}_{i}+\beta_{2} X_{i,t}+\varphi_{k}+u_{i,t}
\end{equation}
where $\hat{F}_{i}$ is the predicted value of fragmentation estimated from Equation 3, $\varphi_{k}$ refers to county-level fixed effects and $u_{i,t}$ is the error term. $X_{i,t}$ is a matrix of control variables including the level and change of foreign-direct investment, unemployment rate, population density, and distance to the closest border.

Results presented in Table 1 confirm that social network fragmentation, instrumented by urban structure indicators, is positively related with income inequality. This result strengthens the validity of the relationship between social network fragmentation and inequality. It also suggests that urban structure is an important indicator of social network outcomes that coincide with inequality. The control variables inform us that densely populated towns have lower levels of inequality than sparsely populated towns and inequality in towns close to the border (equals with periphery in most cases) is above average. The findings are robust in regardless of the three approaches of urban topology and their composite measure. Supporting Information 6 contains complete regression tables of the second stages of the 2SLS model.

Model statistics suggest that urban topology indicators are strong instruments of network fragmentation, confirmed by an F-test of the first stage regression. A Wu-Hausman test confirms that they are not significantly correlated with the second stage dependent variable: income inequality. With the exception of the $SCA$ regression, the instrumental variable models provide better fit than OLS regressions using the original urban structure measures instead of fragmentation as confirmed by a Sargan test. In sum, urban topology indicators are reliable instruments of social network fragmentation for the purposes of our two-stage regression.

\begin{table}[!t] \centering 
  \caption{We estimate the relationship between social network fragmentation and income inequality using the urban topology indices as instruments for fragmentation using Equation 4. We report the second stage of the 2SLS regressions.} 
  %\begin{adjustbox}{max width=.48\textwidth}
\begin{tabular}{@{\extracolsep{\fill}}l c c c c }
\\[-1.8ex]\hline 
\hline \\[-1.8ex] 
 & \multicolumn{4}{c}{Dependent variable: Gini$_{2016}$} \\ 
\hline \\[-1.8ex]
%\cline{2-6} 
\\[-1.8ex] & \multicolumn{4}{c}{$Instrumental$ $Variable$} \\ 
\\[-1.8ex] & \multicolumn{1}{c}{\textit{ADC}} & \multicolumn{1}{c}{\textit{SPB}} & \multicolumn{1}{c}{\textit{SCA}} & \multicolumn{1}{c}{\centering{\textit{CUTI}}}\\ 
\hline \\[-1.8ex] 
 Fragmentation & $0.408^{***}$ & $0.288^{**}$  & $0.533^{***}$ & $0.428^{***}$  \\ 
  & $(0.153)$ & $(0.138)$  & $(0.146)$ & $(0.119)$  \\ 
  & & & &  \\ 
 Population density & $-0.092^{*}$ & $-0.067$  & $-0.118^{**}$ & $-0.096^{**}$  \\ 
  & $(0.055)$ & $(0.053)$  & $(0.052)$ & $(0.048)$  \\ 
  & & & &  \\ 
Distance to border & $-0.243^{***}$ & $-0.254^{***}$ & $-0.231^{***}$ & $-0.241^{***}$ \\ 
 & $(0.059)$ & $(0.058)$  & $(0.062)$ & $(0.062)$ \\ 
  & & & &  \\ 
 Constant & $-0.386$ & $-0.380$  & $-0.392$ & $-0.389$  \\ 
  & $(0.369)$ & $(0.372)$ & $(0.330)$ & $(0.375)$  \\ 
  & & & &  \\ 
\hline \\[-1.8ex] 
County FE & Yes & Yes  & Yes & Yes  \\ 
Controls & Yes & Yes  & Yes & Yes  \\ 
\hline \\[-1.8ex] 
First Stage $F$-test & $22.290^{***}$ & $26.754^{***}$ & $24.009^{***}$ & $33.991^{***}$ \\ 
Wu-Hausman test & $1.107$  & $0.011$ & $3.729$ & $1.848$   \\ 
Sargan test & $0.051$  & $1.400$ & $5.349^{*}$ & $0.136 $  \\ 
\hline \\[-1.8ex] 
Observations & \multicolumn{1}{c}{473} & \multicolumn{1}{c}{473} & \multicolumn{1}{c}{473} & \multicolumn{1}{c}{473} \\ 
R$^{2}$ & \multicolumn{1}{c}{0.231} & \multicolumn{1}{c}{0.245} & \multicolumn{1}{c}{0.192} & \multicolumn{1}{c}{0.226}  \\ 
Adjusted R$^{2}$ & \multicolumn{1}{c}{0.186} & \multicolumn{1}{c}{0.200}  & \multicolumn{1}{c}{0.145} & \multicolumn{1}{c}{0.181} \\ 
Res.St.Err. (df = 446) & \multicolumn{1}{c}{0.902} & \multicolumn{1}{c}{0.894} & \multicolumn{1}{c}{0.924} & \multicolumn{1}{c}{0.905} \\ 
\hline
\hline \\[-1.8ex] 
\textit{Note:}  & \multicolumn{4}{l}{$^{*}$p$<$0.1; $^{**}$p$<$0.05; $^{***}$p$<$0.01} \\ 
\end{tabular}
%\end{adjustbox}
\label{table:Table1}
\end{table}

\section*{Discussion}
In this paper, we have established a new evidence using Big Data that the fragmentation of social network structure is positively associated with income inequality in cities and towns. Moreover, we have found that the relationship is dynamic - fragmentation and existing inequalities predict a significant growth in inequality in the future. The physical arrangement of a city's residential areas, the loci of social interactions, is also connected to social network fragmentation. We observe a tendency: if the urban fabric contains significant distances, physical barriers, or spatially concentrated amenities, social networks tend to be more fragmented.

We cannot prove the following story of cause and effect: that poorly designed cities fragment the social network and amplify economic inequality. Nevertheless our observations give us the confidence to propose that the rise of inequalities in towns may be fruitfully blunted by wise urban planning. We hypothesize that improving access across neighborhoods, facilitating mixing within them, and supporting a more equal distribution of services can cure broken social networks and improve economic outcomes across the board.

\section*{Materials and Methods}
\subsection*{Materials}
Data tenure of the iWiW online social network is controlled by a non-disclosure agreement. The data, besides other information, includes self-reported location of 2.8M users and their social connections reported on the OSN website. Access can be requested in email addressed to: \href{mailto:lengyel.balazs@krtk.mta.hu}{\nolinkurl{lengyel.balazs@krtk.mta.hu}}.

Town-level aggregate information including income distributions, population distribution according to school, age, ethnic and religious groups, population density, unemployment, distance from border and foreign-direct investment was collected from \url{https://www.teir.hu}.

Corine Land Cover (CLC) data of built up residential areas including continuous and discontinuous urban fabric according to CLC 2012 were collected from the \url{https://land.copernicus.eu/pan-european/corine-land-cover} website.

Geographic data on the location of residential areas, rivers, railroads, and major roads was collected from \url{https://data2.openstreetmap.hu/hatarok/}.

Data on POI listed as "amenities" was downloaded using the \url{https://download.geofabrik.de/} website.

\subsection*{Methods}
To calculate $ADC$, we randomly located points on the polygons of residential zones in the CLC database. The number of points in each town refers to it's total population and, because we aim to create the measure reflecting on urban topology, the number of points in a polygon is a function of the polygon's area. 
Based on this randomized spatial distribution, we estimated the center of gravity for the town and calculated the average distance of points from it, following:
\begin{equation}
ADC_{i} = \frac{\sum_{p}^{P}D_{p,c}}{P} / S_{i} 
\end{equation}
where $D_{p,c}$ denotes the distance of points $p$ to the estimated center of gravity $c$ out of $P$ points and $S_{i}$ refers to the size of the town's area. The value of $ADC$ is small for a compact settlements and is large in towns with remote population fragments. 

To calculate $SPB$, we adapt the measure of the physical division of the residential areas of cities known as the Railroad Division Index \cite{ananat2011wrong}. We source data on residential-zoned areas in each settlement in the OSM dataset and cut the polygons by the rivers, major roads, and railroads in the settlement. Then, we calculate the resulting dispersion of its residential area across disconnected components:
\begin{equation}
SPB_{i} = 1 - \sum_{a}(S_{a}/S_{i}) ^{2}
\end{equation}
where $S_{i}$ refers to the size of the town's area and $S_{a}$ denotes size of area $a$ after barrier dispersion. The value of $SPB$ is small for settlements that are not divided by barriers and large for those where barriers disconnect large fractions of residential areas. 

To measure how much spatially concentrated the amenities are in the town, we identify spatial clusters of POI by applying a DBSCAN algorithm with 500 meters radius. This algorithm groups those amenities together that are in the close neighborhood of each other. The spatial concentration of the groups is then quantified with the probabilistic entropy of the size distribution of spatial clusters multiplied by minus $1$:
\begin{equation}
SCA_{i} = \frac{ \sum_{c}(p_{c} \times \log p_{c})}{n(c)} / S_{i}
\end{equation}
where $c$ refers to spatial clusters and $p_{c}$ is the number of POIs in $c$ over the total number of clusters in the town $n(c)$ . The value of $SCA$ is high for settlements where amenities are concentrated in few spatial clusters and small for those where amenities are evenly.

\section*{Acknowledgements} B.S., J.K., \'A.J, and B.L. acknowledge financial support received from National Office for Researcher and Innovation (OTKA K129124). R.D.C. as Newton International Fellow of the Royal Society acknowledges support from the Royal Society, the British Academy, and the Academy of Medical Sciences (Newton International Fellowship, NF170505)

\section*{Author Contributions} J.W., B.S., J.K., B.L. designed the research, G.T., J.W., \'A.J., R.D.C  conceived the experiments, G.T., J.W. and B.L. analyzed the results. All authors wrote and reviewed the manuscript.

\bibliographystyle{spphys}       
\bibliography{references}

\clearpage

\section*{Supplementary Information}
\section*{Supporting Information 1: Gini coefficient as the measure of local income inequalities}

We adopt the Gini index to quantify economic inequality from income distributions. This widely used indicator is defined by the following equation:
\begin{equation}
G_{i,t}=\frac{\sum_{p=1}^{n}\sum_{q=1}^{n} \mid x_{p} - x_{q} \mid }{2n\sum_{p=1}^{n}  x_{q}}
\end{equation}
where $i$ refers to a town, $t$ denotes the year (either 2011 or 2016), $x_{p}$ and $x_{q}$ are the sum of total income in income categories $p$ and $q$, and \textit{n} denotes the number of income categories within towns.

\section*{Supporting Information 2: Online social network data and representativity}

\subsection*{About the data}

The iWiW (International Who Is Who) was launched in 2002 and shortly became the most widely used online social network in Hungary. At its peak, it was one of the most visited national websites, where approximately 30 per cent of the Hungarian population over 14 years registered as of January 2013, and covering the majority of internet users of the country. During the first few years of operation iWiW provided only basic functionalities, mostly built around finding present and former friends, classmates, colleagues, and looking through one's acquaintance's' acquaintances. Later, photo upload, news-feed (similar to Facebook), messaging, applet to visualize connections and the ability to develop external applications was introduced to the service. But all these came too late, so due to the increasing maintenance costs, low profitability and tough competition from Facebook the site was closed down permanently on June 30, 2014. 
Although the number of daily visitors begun to fall back significantly from 2011-2012, users rarely deleted their profiles: they just abandoned the service. 

In February 2013, the entire dataset of iWiW with basic user information (i.e. date of registration, gender, age, etc.) and connection data (establishment of friendship ties) was exported for scientific research purposes. 

\subsection*{Representativity}
\begin{figure*} 
\centering
\includegraphics[width=\linewidth]{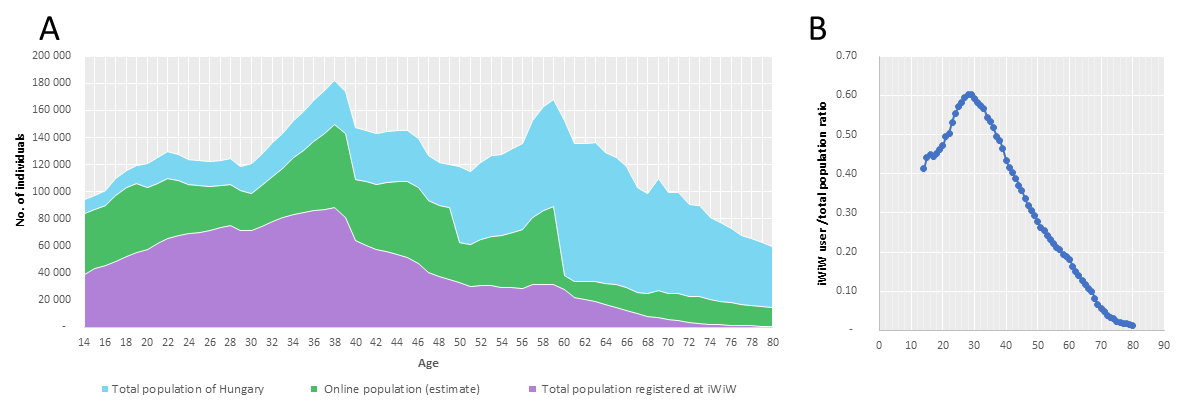}
\caption{\textbf{iWiW users compared to the Hungarian population} 
\textbf{(A)} Comparison of (1) total population of Hungary, (2) estimated online population using nationally representative survey data from 2013, (3) total population registered at iWiW.\textbf{(B)} Ratio of iWiW users and total population of Hungary by age.}
\label{fig:Fig.1}
\end{figure*}

Never before was such a large scale dataset available for research regarding the social connections of the Hungarian population. Use of the service was limited to those aged over 14, so theoretically the maximum number of potential users was 8,2 million people in Hungary. The total number of users who chose a Hungarian settlement as their home location reached more than 4 million by early 2013 (another 600.000 users were outside Hungary). This implies that about 50\% of Hungarians were part of the network. Considering the level of internet users measured by nationally representative surveys (76\%) in 2013, around two-third of the online population were iWiW users. 
In our analysis, social connections represented online are used as a proxy of real-life social connections. This approach is certainly a simplification of the complex social reality, but we argue that despite our data is imperfect and has certain limitations (i.e. we do not know about the nature of social connections, their strength, frequency of communication, etc.), until now it is still the best available source, and there is no such systematic bias in the data that would question the validity of the analysis. 
The latter is demonstrated in Figure \ref{fig:Fig.1}(A) comparing the number of individuals by age (14 to 80) for the total population of Hungary, its estimated online population, and the number of users registered at iWiW. Until 60 years the representation of iWiW users follows the estimated number of internet users without any serious deviation. (B) The iWiW user/total population ratio reaches its maximum (~60\%) around the age of 30, and then starts decreasing continuously, and falling below 30 \% above 50, and 10\% above 70 years. However, the economically active population of Hungary was well represented on the network. 

\subsection*{Potential biases in geographical representativity}
\begin{figure}  
\centering
\includegraphics[width=0.7\textwidth]{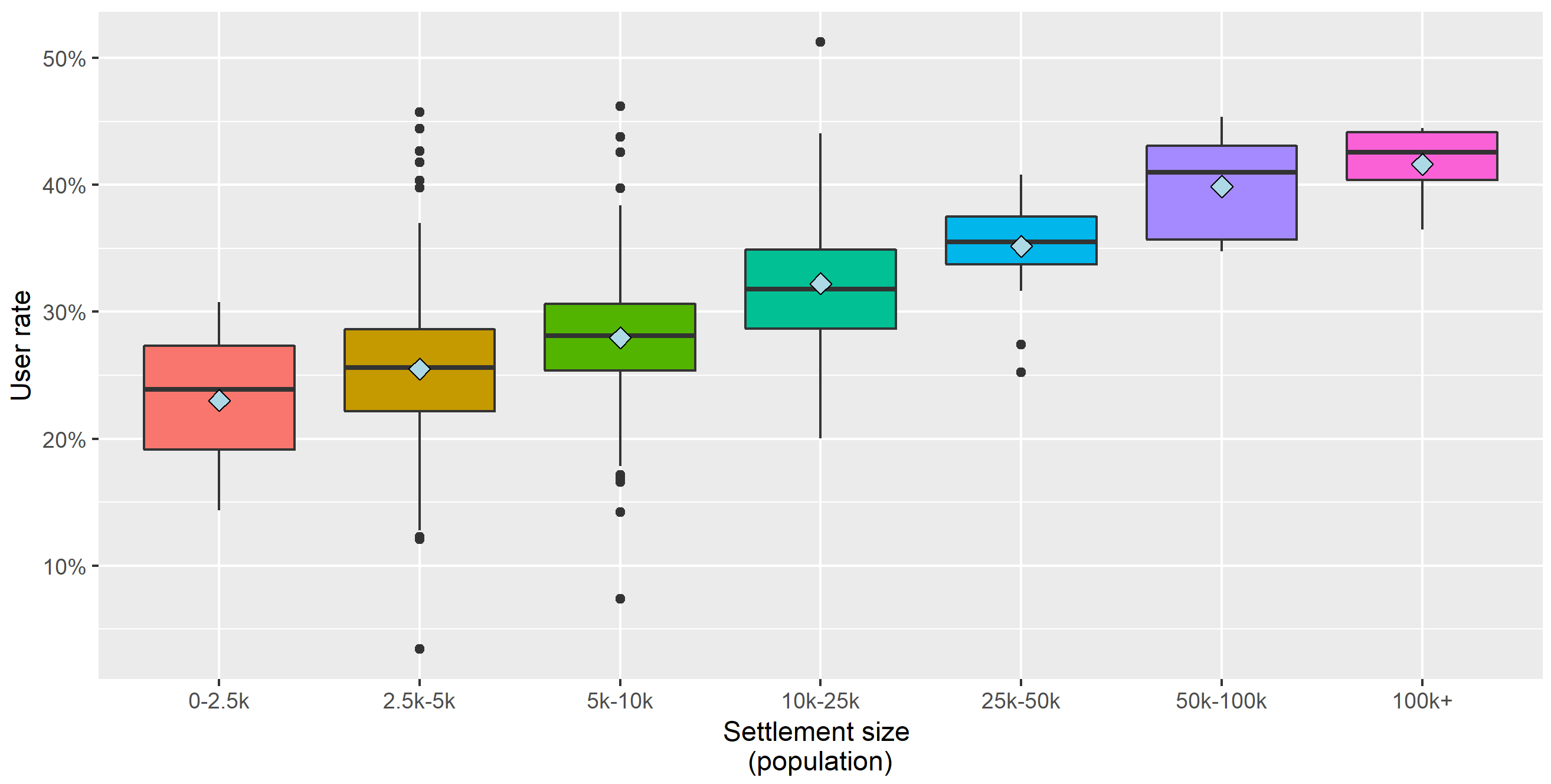}
\caption{iWiW user rate by settlement size categories}
\label{Fig.2}
\end{figure}

\begin{figure}  
\centering
\includegraphics[width=1\textwidth]{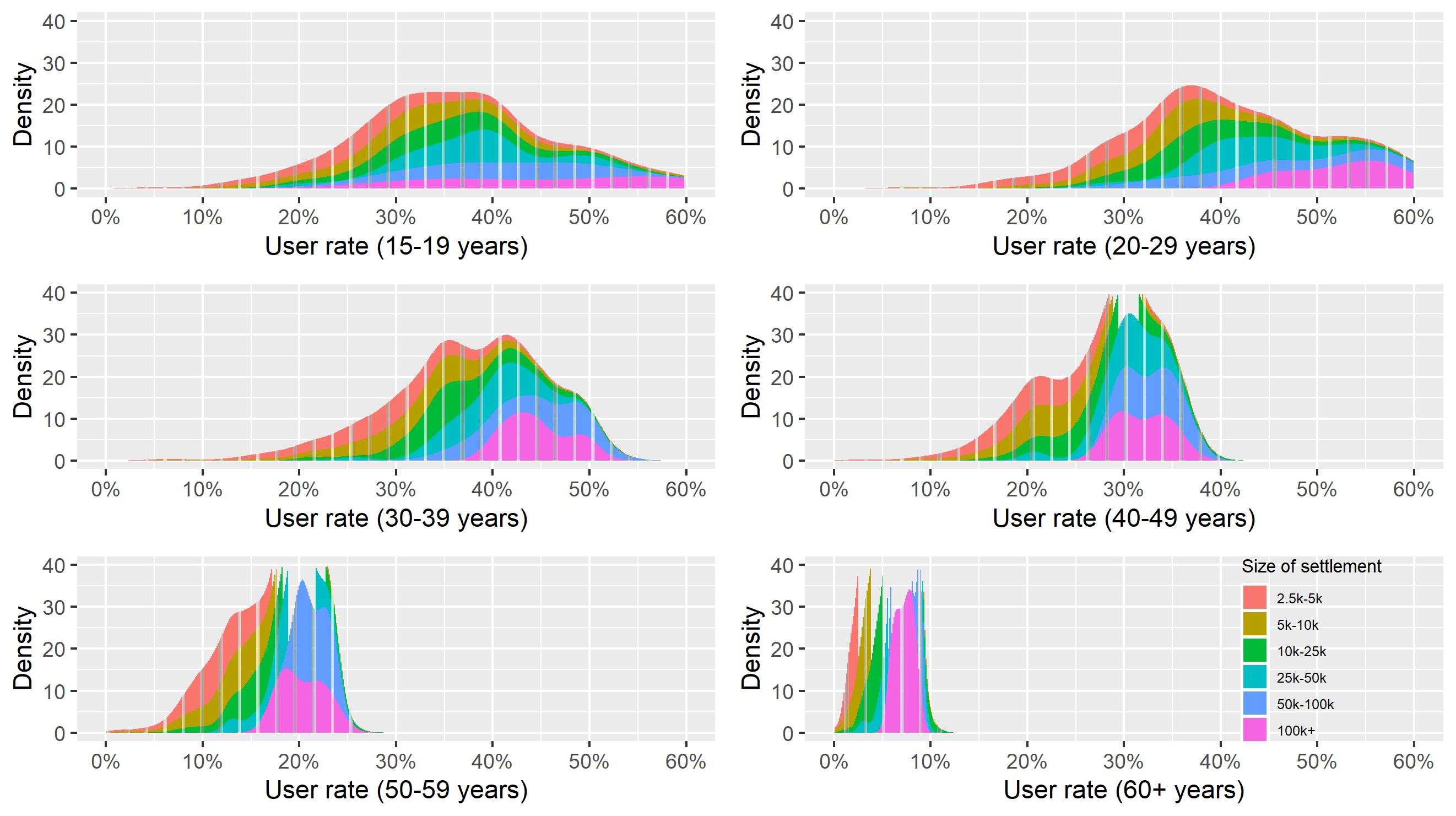}
\caption{Density function of iWiW users in different age groups by size of settlement}
\label{Fig.3}
\end{figure}

Since our analysis is focusing on individual social connections aggregated at settlement level, it is necessary to check for under-representation of certain types of settlements according to their size. The diffusion of innovations - such as the use of an online social network - follows more or less universal patterns, where age, level of education and location play a crucial role. During the fist few years of its life cycle iWiW was mostly used by young, highly educated urban population. Later, more and more elderly people joined from rural areas of the country, however their level of penetration never reached that of the former groups. \ref{Fig.2} and \ref{Fig.3} illustrate that the overall level of iWiW user rate by settlement size varied between 23\% (for small villages) and 42\% (for major cities) . Since elderly, lower educated people are over-represented in smaller settlements these figures are not suprising. However, the number of outlier settlements is relatively low, legitimating the use of our data.  

\newpage

\section*{Supporting Information 3: Principal components of urban topology indicators}

To decrease the dimensions of the urban topology approaches taken in the main text, we constructed the \textit{Composite Urban Topology Index} ($CUTI$)from \textit{Average Distance from the Center} ($ADC$), Segregation by \textit{Physical Barriers} ($SPB$), and \textit{Spatial Concentration of Amenities} ($SCA$) by using principal component analysis. Because we cannot argue against reverse causality in the case of $SCA$, as a robustness check, we constructed the Principal Component from $ADC$ and $SPB$ indices only $PC(ADC,SPB)$. Both of the composite measures have been tested in the remaining empirical analysis and results are reported on both in Supporting Information 5 and 6. In the main text, we report results from using the $CUTI$ only. 

\begin{table}[ht]
 \caption{The components of urban topology} 
\centering
\begin{tabular}{rcc}
  \hline
 & \textit{PC(ADC, SPB)}  & \textit{Composite Urban Topology Index}   \\ 
  \hline
  \textit{Average Distance from the Center} (ADC) & $0.710$ & $0.612$  \\ 
  \textit{Segregation by Physical Barriers} (SPB) & $.0710$  & $0.612$  \\ 
  \textit{Spatial Concentration of Amenities} (SCA) &  & $0.495$  \\ 
  \hline
  \hline
  Eigenvector value of the first component & $1.4$ & $1.60$ \\ 
  Variance explained by the first component & $.70$ & $.55$ \\ 
   \hline

\end{tabular}
\end{table}

Table S1 summarizes information about the principal component analysis. \textit{Composite Urban Topology Index}: 56 percentage of the variance in all three measures is explained by the first component of the principal component analysis using $ADC$, $SPB$, and $SCA$ with the eigenvector value of 1.66. \textit{PC(ADC, SPB)}: 70 percentage of the variance in both measures is explained by the first component of the principal component analysis with the eigenvector value of 1.4. 

High values of all urban topology approaches refer to high spatial segregation induced by distance ($ADC$), physical barriers ($SPB$) and concentration of amenities ($SCA$). Consequently, high levels of both \textit{PC(ADC, SPB)} and the \textit{CUTI} refers to high spatial segregation in all dimensions included.

\newpage

\section*{Supporting Information 4: Controllers description, variable distribution and correlation}

To evaluate the importance of urban structure in social network fragmentation in towns, we apply a machine learning approach and consider further social and demographic factors that can be sources of social segregation besides urban topology. Description of these variables are as follows:
\begin{itemize}
  \itemsep0em
  \item \textit{Ethnic fragmentation}: We collected data of population distribution across ethnic groups (hungarian, roma, german etc.) from \url{https://www.teir.hu} and calculated the entropy of the size distribution of ethnic groups. This indicator is high if ethnic groups in the town have similar sizes. Because link formation is less likely across groups of similar size than between a small group and a large group \cite{currarini2010identifying}, we expect a positive correlation between the index and social network fragmentation.
  \item \textit{Religious fragmentation}: We collected data of population distribution across confession groups (catholic, lutheran, muslim etc.) from \url{https://www.teir.hu} and calculated the entropy of the size distribution. The indicator is high if religious groups in towns have similar sizes. Like in the case of Ethnic fragmentation, we expect a positive correlation between the index and social network fragmentation.
  \item \textit{Political fragmentation}: We calculate the coefficient of variance in the vote share given to right-wing across voting districts in the town. The indicator is high if voting district differ in terms of political preferences. In our specific country case, there is a large ideological difference between the governing right-wing party and the opponent parties, which might be reflected in everyday social interactions as well. Therefore, we expect a positive correlation between the index and social network fragmentation. Data on parliamentary elections in Hungary was collected directly from the Hungarian National Election Office's official website: \url{https://www.valasztas.hu/}. Voting outcomes for the different party lists are available at the level of voting precincts.
 \item \textit{Education inequalities}: We calculate the coefficient of variance in 6th grade math exam. The indicator is high if there is large differences across primary schools in the commuting zone of the town. The quality of schools plays an important role in opportunities for individual progress. Further, school quality differences reflect the divergence of human capital accumulation in the town across generations. Consequently, we expect a positive correlation between this indicator and social network fragmentation. Education data was collected from the national 6th grade competence test in mathematics, that includes individual level data on all primary school students in Hungary in 2011. The raw data is available from the Databank of the Research Centre for Economic and Regional Studies, Hungarian Academy of Sciences. Access can be requested at \url{http://www.krtk.mta.hu/szervezet/adatbank/}.\end{itemize}

Figure S4 illustrates the distribution of the above social segregation variables and our urban topology indicators and their correlation.

In the second stage of the 2SLS regression (Equation 4 in the main text), we include the following control variables that are expected to influence inequalities in towns:
\begin{itemize}
  \itemsep0em
  \item \textit{Population density}: number of inhabitants divided by the size of the residential area.
   \item \textit{High school}: the ratio of residents with high school degree or above.
  \item \textit{Age}: the ratio of residents older than 60 years.
 \item \textit{Unemployment ratio}: number of unemployed people as a percentage of labour force.
  \item \textit{Distance to border}: the distance from the nearest border measured in kilometers.
  \item \textit{Foreign investment}: revenue capital owned by foreign firms, measured in 1000 Hungarian Forint.
\end{itemize}

All data required to calculate the control variables was retrieved from  \url{https://www.teir.hu}. Figure S5 illustrate distribution of control variables and social network fragmentation, and their correlation.

\newpage

\begin{figure*}[!h]
\centering
\includegraphics[width=0.7\linewidth]{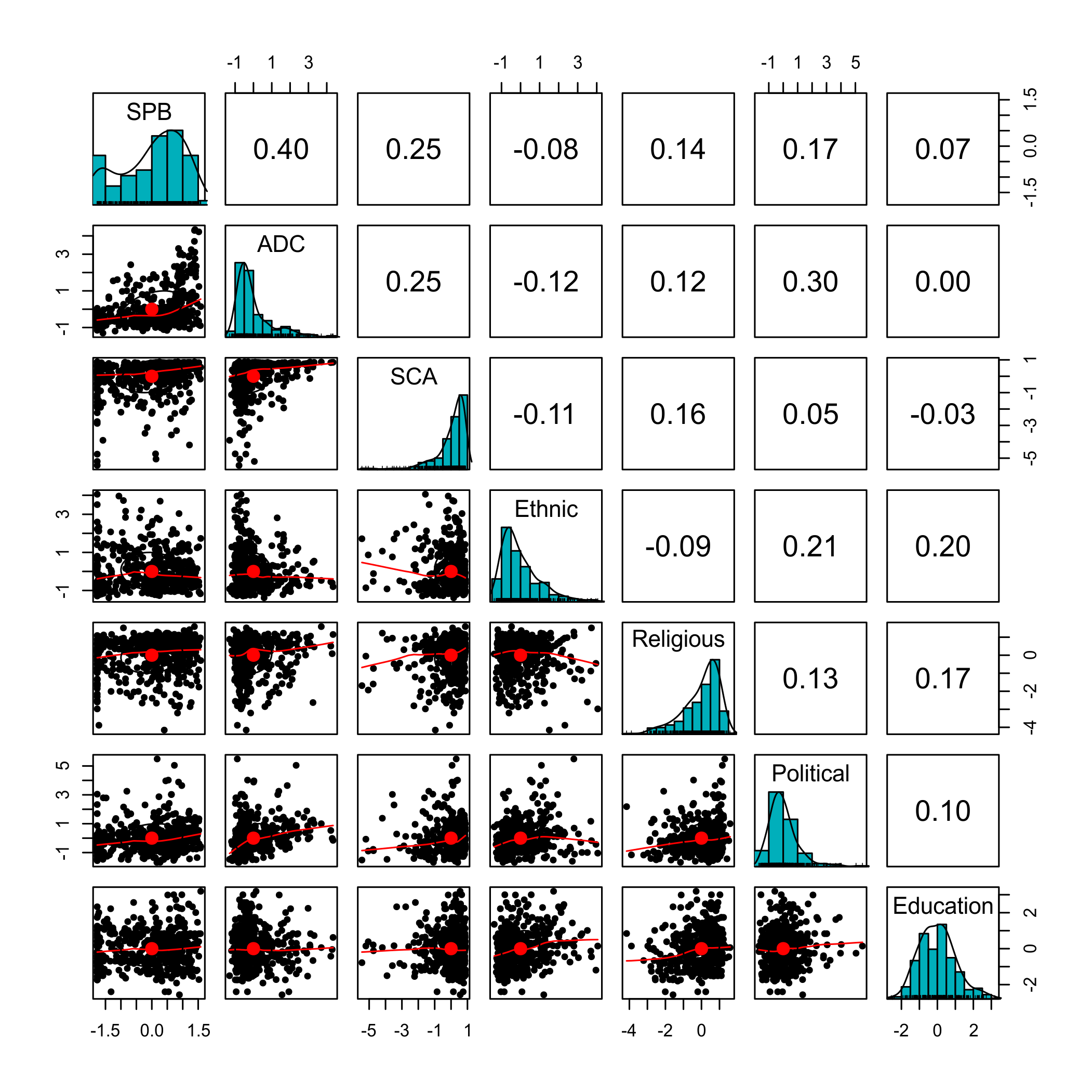} 
\caption{Correlation and distribution of the variables in the Random Forest exercise. All the variables are standardized into z-scores.}
\label{fig:subim1}
\end{figure*}

\newpage

\begin{figure*}[!h]
\centering
\includegraphics[width=0.8\linewidth]{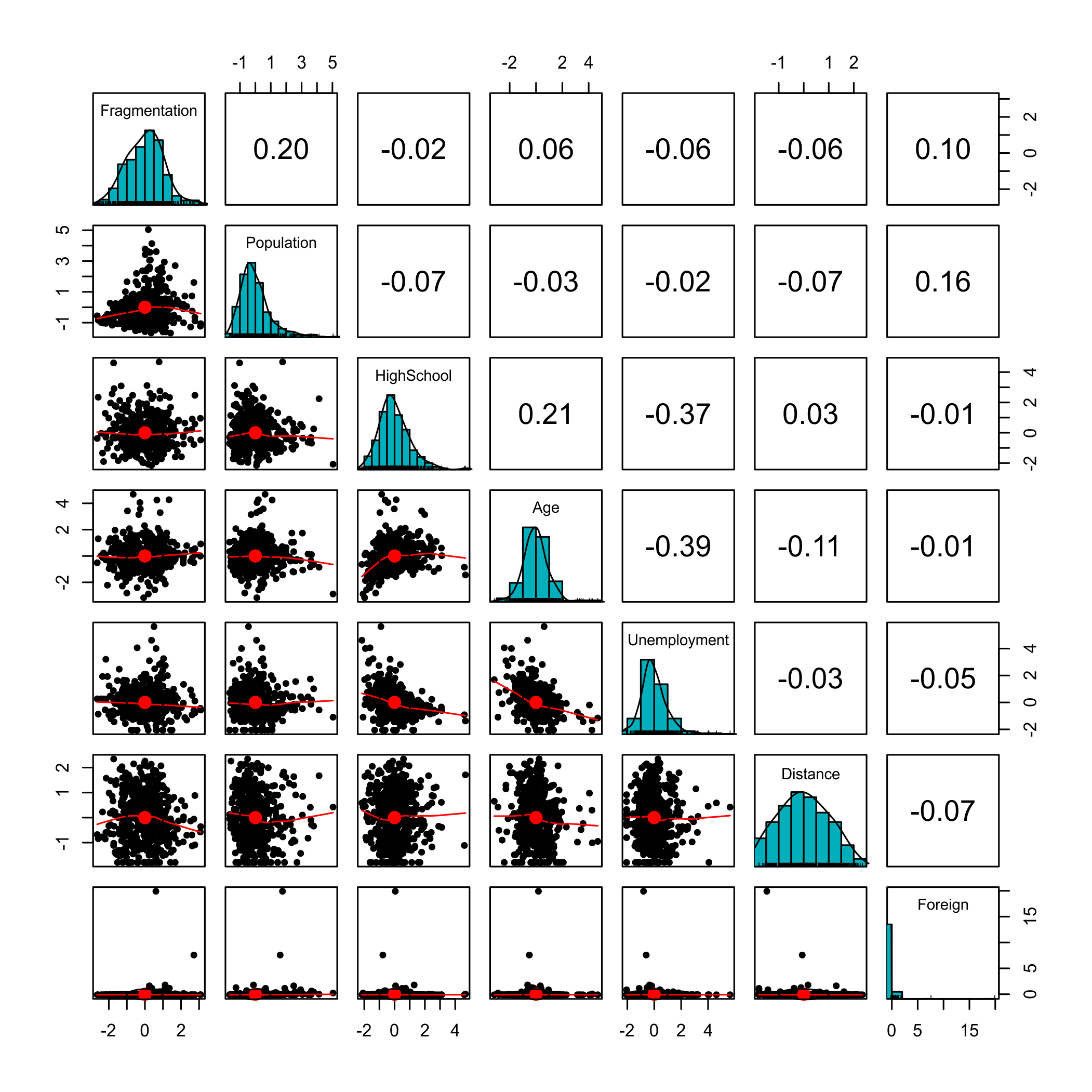}
\caption{Correlation and distribution of the variables in the second stage of the 2SLS regression. All the variables are standardized into z-scores.}
\label{fig:subim2}
\end{figure*}

\newpage

\section*{Supporting Information 5: The importance of urban structure in network fragmentation compared to other dimensions of segregation}

\begin{figure*}[!h]  
\centering
\includegraphics[width=\linewidth]{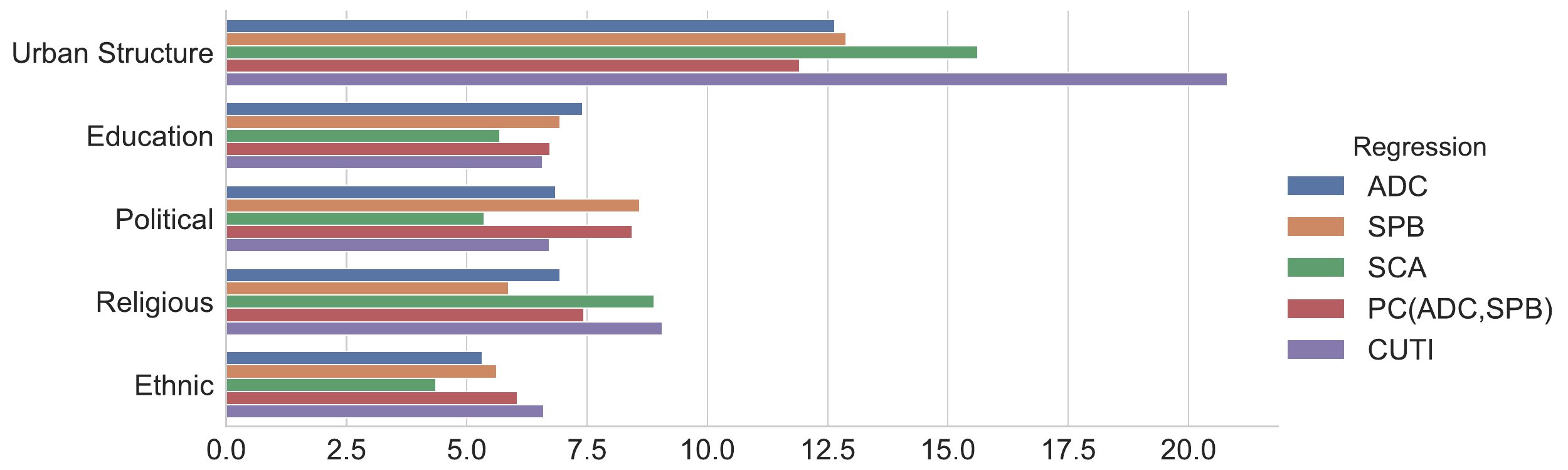}
\caption{\textbf{Variable importance from the Random Forest prediction of social network fragmentation in towns.} Regardless of taking different approaches, urban topology outperforms other determinants of social segregation.}
\label{fig:Fig.4}
\end{figure*}

We apply a Random Forest technique to rank the drivers of social network fragmentation in towns. We estimate $F_{i}$ by randomly combining urban topology indicators and alternative determinants of fragmentation in 500 regressions based on decision trees. To predict variable importance we take a random sample from the decision trees and calculate the mean squared error ($MSE$) of the predictions applying the formula $\sum_{1}^{e}\frac{(F_{i}-\hat{F}_{i})^2}{e}$. To quantify the importance of each determinant, we let the value of the variable in focus randomly shuffle around its mean while keeping other variables in the regression fixed and re-calculate $MSE$. Applying this technique informs us about the importance of observed values of the variables in focus compared to a randomized distribution \cite{gromping2009variable}. Results illustrated in Figure \ref{fig:Fig.4} confirm that every aspect of urban structure outperforms the alternative drivers in predicting network fragmentation.

\newpage

\section*{Supporting Information 6: Full tables of the 2SLS models}

\begin{table}[!htbp] \centering 
  \caption{Fragmentation estimates, 2SLS regression, first stage} 
  \label{tab2} 
\begin{tabular}{@{\extracolsep{5pt}}lc c c c c  } 
\\[-1.8ex]\hline 
\hline \\[-1.8ex] 
 & \multicolumn{5}{c}{Dependent variable:$Fragmentation_{i}$} \\ 
\cline{2-6} 
\\[-1.8ex] & \multicolumn{5}{c}{} \\ 
\\[-1.8ex] & \multicolumn{1}{c}{(1)} & \multicolumn{1}{c}{(2)} & \multicolumn{1}{c}{(3)} & \multicolumn{1}{c}{(4)} & \multicolumn{1}{c}{(5)}\\ 
\hline \\[-1.8ex] 
\textit{SPB}  & $0.168^{***}$ &  &  &  &  \\ 
  & $(0.044)$ &  &  &  &  \\ 
  & & & & & \\ 
\textit{ADC} &  & $0.091^{**}$ &  &  &  \\ 
  &  & $(0.045)$ &  &  &  \\ 
  & & & & & \\ 
\textit{SCA} &  &  & $0.110^{**}$ &  &  \\ 
  &  &  & $(0.046)$ &  &  \\ 
  & & & & & \\ 
\textit{ADC \& SPB} &  &  &  & $0.139^{***}$ &  \\ 
  &  &  &  & $(0.039)$ &  \\ 
  & & & & & \\ 
\textit{Composite Urban Topology Index} &  &  &  &  & $0.153^{***}$ \\ 
  &  &  &  &  & $(0.037)$ \\ 
  & & & & & \\ 
User rate & $0.344^{***}$ & $0.367^{***}$ & $0.355^{***}$ & $0.332^{***}$ & $0.304^{***}$ \\ 
  & $(0.044)$ & $(0.045)$ & $(0.046)$ & $(0.046)$ & $(0.048)$ \\ 
  & & & & & \\ 
 Constant & $-0.000$ & $-0.000$ & $-0.000$ & $0.000$ & $0.000$ \\ 
  & $(0.042)$ & $(0.042)$ & $(0.042)$ & $(0.042)$ & $(0.042)$ \\ 
  & & & & & \\ 
\hline \\[-1.8ex] 
Observations & \multicolumn{1}{c}{473} & \multicolumn{1}{c}{473} & \multicolumn{1}{c}{473} & \multicolumn{1}{c}{473} & \multicolumn{1}{c}{473} \\ 
R$^{2}$ & \multicolumn{1}{c}{0.185} & \multicolumn{1}{c}{0.167} & \multicolumn{1}{c}{0.170} & \multicolumn{1}{c}{0.182} & \multicolumn{1}{c}{0.188} \\ 
Adjusted R$^{2}$ & \multicolumn{1}{c}{0.181} & \multicolumn{1}{c}{0.163} & \multicolumn{1}{c}{0.166} & \multicolumn{1}{c}{0.179} & \multicolumn{1}{c}{0.185} \\ 
Residual Std. Error (df = 470) & \multicolumn{1}{c}{0.905} & \multicolumn{1}{c}{0.915} & \multicolumn{1}{c}{0.913} & \multicolumn{1}{c}{0.906} & \multicolumn{1}{c}{0.903} \\ 
F Statistic (df = 2; 470) & \multicolumn{1}{c}{$53.259^{***}$} & \multicolumn{1}{c}{$47.082^{***}$} & \multicolumn{1}{c}{$48.085^{***}$} & \multicolumn{1}{c}{$52.338^{***}$} & \multicolumn{1}{c}{$54.501^{***}$} \\ 
\hline 
\hline \\[-1.8ex] 
\textit{Note:}  & \multicolumn{5}{r}{$^{*}p<0.1$; $^{**}p<0.05$; $^{***}p<0.01$} \\ 
\end{tabular} 
\end{table} 

\newpage
\begin{table}[!htbp] \centering 
  \caption{Inequality estimates, 2SLS regression, second stage} 
  \label{tab:1} 
\begin{tabular}{@{\extracolsep{5pt}}lc c c c c  } 
\\[-1.8ex]\hline 
\hline \\[-1.8ex] 
 & \multicolumn{5}{c}{\textit{Dependent variable: $Gini_{2016}$}} \\ 
\cline{2-6} 
\\[-1.8ex] & \multicolumn{5}{c}{$Instrumental$ $Variable$} \\ 
\\[-1.8ex] & \multicolumn{1}{c}{\textit{SPB}} & \multicolumn{1}{c}{\textit{ADC}} & \multicolumn{1}{c}{\textit{SCA}} & \multicolumn{1}{c}{\textit{PC(ADC,SPB)}} & \multicolumn{1}{p{2cm}}{\centering{\textit{Composite Urban Topology Index}}}\\ 
\hline \\[-1.8ex] 
Fragmentation & $0.288^{**}$ & $0.408^{***}$ & $0.533^{***}$ & $0.338^{**}$ & $0.428^{***}$ \\ 
  & $(0.138)$ & $(0.153)$ & $(0.146)$ & $(0.138)$ & $(0.119)$ \\ 
  & & & & & \\ 
Population density & $-0.067$ & $-0.092^{*}$ & $-0.118^{**}$ & $-0.077$ & $-0.096^{**}$ \\ 
  & $(0.053)$ & $(0.055)$ & $(0.052)$ & $(0.052)$ & $(0.048)$ \\ 
  & & & & & \\ 
High school & $0.001$ & $0.001$ & $0.001$ & $0.001$ & $0.001$ \\ 
  & $(0.004)$ & $(0.004)$ & $(0.004)$ & $(0.004)$ & $(0.004)$ \\ 
  & & & & & \\ 
 Age & $0.687$ & $0.582$ & $0.473$ & $0.643$ & $0.564$ \\ 
  & $(0.824)$ & $(0.833)$ & $(0.854)$ & $(0.826)$ & $(0.834)$ \\ 
  & & & & & \\ 
 Unemployment ratio & $-1.221$ & $-1.043$ & $-0.856$ & $-1.147$ & $-1.012$ \\ 
  & $(1.558)$ & $(1.575)$ & $(1.614)$ & $(1.561)$ & $(1.576)$ \\ 
  & & & & & \\ 
Distance to border & $-0.254^{***}$ & $-0.243^{***}$ & $-0.232^{***}$ & $-0.249^{***}$ & $-0.241^{***}$ \\ 
  & $(0.058)$ & $(0.059)$ & $(0.060)$ & $(0.058)$ & $(0.062)$ \\ 
  & & & & & \\ 
Foreign investment & $0.075^{*}$ & $0.070$ & $0.064$ & $0.073^{*}$ & $0.069$ \\ 
  & $(0.044)$ & $(0.045)$ & $(0.046)$ & $(0.044)$ & $(0.045)$ \\ 
  & & & & & \\ 
$\Delta$ Foreign investment$_{2011-2016}$ & $-0.009$ & $-0.012$ & $-0.015$ & $-0.010$ & $-0.013$ \\ 
  & $(0.016)$ & $(0.017)$ & $(0.017)$ & $(0.016)$ & $(0.016)$ \\ 
  & & & & & \\ 
 Constant & $-0.380$ & $-0.386$ & $-0.392$ & $-0.382$ & $-0.387$ \\ 
  & $(0.369)$ & $(0.372)$ & $(0.330)$ & $(0.370)$ & $(0.373)$ \\ 
  & & & & & \\ 
\hline \\[-1.8ex] 
County FE & Yes & Yes  & Yes & Yes & Yes \\ 
\hline \\[-1.8ex] 
Weak instruments & $26.754^{***}$ & $22.290^{***}$ & $24.009^{***}$ & $9.635^{***}$ & $33.991^{***}$ \\ 
Wu-Hausman tests  & $0.011$ & $1.107$  & $3.729$ & $0.275$ & $1.848$  \\ 
Sargan tests & $1.400$ & $0.051$  & $5.349 ^{*}$ & $0.373$ & $0.136$ \\ 
\hline \\[-1.8ex] 
Observations & \multicolumn{1}{c}{473} & \multicolumn{1}{c}{473} & \multicolumn{1}{c}{473} & \multicolumn{1}{c}{473} & \multicolumn{1}{c}{473} \\ 
R$^{2}$ & \multicolumn{1}{c}{0.245} & \multicolumn{1}{c}{0.231} & \multicolumn{1}{c}{0.192} & \multicolumn{1}{c}{0.242} & \multicolumn{1}{c}{0.226} \\ 
Adjusted R$^{2}$ & \multicolumn{1}{c}{0.200} & \multicolumn{1}{c}{0.186} & \multicolumn{1}{c}{0.145} & \multicolumn{1}{c}{0.197} & \multicolumn{1}{c}{0.181} \\ 
Residual Std. Error (df = 446) & \multicolumn{1}{c}{0.894} & \multicolumn{1}{c}{0.902} & \multicolumn{1}{c}{0.924} & \multicolumn{1}{c}{0.896} & \multicolumn{1}{c}{0.905} \\ 
\hline 
\hline \\[-1.8ex] 
\textit{Note:}  & \multicolumn{5}{r}{$^{*}p<0.1$; $^{**}p<0.05$; $^{***}p<0.01$} \\ 
\end{tabular} 
\end{table}

\end{document}